\documentclass[pdflatex,sn-mathphys-num]{sn-jnl}

\usepackage{graphicx}%
\usepackage{multirow}%
\usepackage{amsmath,amssymb,amsfonts}%
\usepackage{amsthm}%
\usepackage{mathrsfs}%
\usepackage[title]{appendix}%
\usepackage{xcolor}%
\usepackage{textcomp}%
\usepackage{manyfoot}%
\usepackage{booktabs}%
\usepackage{algorithm}%
\usepackage{algorithmicx}%
\usepackage{algpseudocode}%
\usepackage{listings}%
\usepackage{lineno}%
\usepackage[capitalise]{cleveref}%

\begin{document}

\title[Sounding Instrument Localisation]{Sounding Surgical Instrument Localisation through Audio-Visual Correspondence}

 
\author[1,2]{\fnm{Dominic} \sur{Behling}}

\author[1]{\fnm{Jonas} \sur{Hein}}

\author*[1]{\fnm{Matthias} \sur{Seibold}}\email{matthias.seibold@balgrist.ch}

\affil[1]{\orgdiv{Research in Orthopedic Computer Science (ROCS), Balgrist University Hospital, University of Zurich, Zurich, Switzerland}}
\affil[2]{\orgname{Eberhard Karls Universität Tübingen}, \country{Germany}}


\abstract{
\textbf{Purpose:} 
Surgical scene understanding is a prerequisite for context-aware and intelligent systems in the operating room. Current methods rely mostly on visual data, although sound carries complementary information on surgical activity. This work investigates whether the active instrument can be localised from a single microphone and a camera by learning the correspondence between the appearance and sound of surgical instruments.
\\
\textbf{Methods:} 
We adapt weakly supervised audio-visual localisation, which learns from the co-occurrence of image and sound, to the surgical setting. Trained without any spatial annotation, the model predicts a heatmap over the sounding instrument. We extend its training objective by silence and area terms, which suppress the output for silence and background noise, and a composite term, which trains the model to select the sounding one of two visible instruments. For training, we record a new corpus of simulated surgical sawing, drilling, and chiselling.
\\
\textbf{Results:} 
On an independent corpus recorded with a different camera and microphone and with optically tracked instrument positions, the model localises the active instrument, separates instrument activity from silence and background noise, and attributes sound correctly when two instruments are present. Ablations show that the silence and area terms are required for detecting activity, and the composite term for selecting between instruments.
\\
\textbf{Conclusion:} 
Audio-visual correspondence learned from in-domain recordings connects the sound of a surgical action to the instrument that produces it. This proof of concept introduces a novel paradigm that enables the detection and localisation of surgical activity without annotating instrument positions.
}
\keywords{Surgical Scene Understanding, Sound Source Localisation, Audio-visual Correspondence, Weakly Supervised Learning, Multimodal Data}



\maketitle
\section{Introduction}\label{sec:introduction}

Over the past decades, Computer Assisted Surgery (CAS) has evolved from the first image-guided navigation systems into a broad technological spectrum that aims at improving the accuracy, safety, and reproducibility of surgical procedures~\cite{Mezger2013}. 
A central building block towards the next generation of intelligent systems in the operating room (OR) is a fine-grained understanding of the surgical scene, its workflow, and the actions performed within it~\cite{hein2017sds}. 
The community has addressed this need, e.g., through automated surgical phase recognition from video~\cite{twinanda2017endonet} and semantic scene graphs that model the interactions between staff, patient, and objects~\cite{oeszoy2022ssg}. 
More recently, surgical digital twins have been proposed as interpretable, geometry-aware representations of the surgical scene that can serve as an internal model for robots and agentic systems~\cite{ding2024dt}.

Current approaches for surgical scene understanding, however, rely almost exclusively on the visual modality. 
Vision-based pipelines require free line-of-sight and are therefore sensitive to occlusions by staff, drapes, and equipment, and they cannot directly capture the physical signature of instrument-tissue interaction, e.g., whether a power tool is running and in contact with the anatomy, which is straightforward to perceive in the acoustic domain. 
Acoustic sensing has consequently been shown to complement vision in a variety of interventional applications, such as tissue characterisation during needle insertion~\cite{illanes2018novel}, error prevention in orthopaedic drilling~\cite{seibold2021realtime}, implant insertion monitoring~\cite{seibold2021femoralstem}, and event detection in the OR~\cite{fuchtmann2024audio}. 
While these works demonstrate that surgical sound carries rich semantic information about \emph{what} is happening and \emph{when}, they do not reveal \emph{where} in the surgical field an action is taking place.

To bridge this gap, sound source localisation (SSL) based on acoustic beamforming with a phased microphone array has recently been introduced to surgery to localise active surgical tools and to fuse acoustic activity maps into dynamic 3D representations of the surgical scene~\cite{seibold2024ssl,hein2026sound}. 
Although these works demonstrated the potential of spatial acoustic information for surgical scene understanding, beamforming-based SSL has considerable limitations that hinder its clinical translation: microphone arrays are bulky and expensive and the computation of acoustic heatmaps is demanding and limits real-time capabilities. 
Moreover, as beamforming localises the \emph{physical} sound source, only a part of the instrument is detected (e.g., the engine of a power drill or the point of hammer impact during chiselling), and the resulting maps are frequently corrupted by artefacts such as acoustic reflections. 
Finally, detection and localisation are solved in separate stages~\cite{hein2026sound}, rather than jointly.

A promising alternative is to exploit the natural co-occurrence of sound and vision. 
Audio-visual sound source localisation (AV-SSL) methods learn to associate sounds with the visual objects that produce them. 
Since audio-visual correspondence arises naturally in video, these methods can be trained in a self- or weakly supervised manner without costly spatial annotations. 
In the general computer vision domain, large-scale benchmarks such as VGG-SS~\cite{chen2021localizing} have driven the development of a variety of approaches based on contrastive learning~\cite{chen2021localizing,mo2022closer,park2026hearing}.
For surgery, this paradigm has the potential to overcome the limitations of beamforming-based SSL: the hardware is reduced to a single consumer-grade camera with a microphone, inference requires only a single forward pass of a neural network, and the full sounding object rather than only its physical emission point is localised, while detection and localisation are addressed jointly. 
To date, however, AV-SSL has not been investigated in the surgical domain. 
In addition, the surgical setting introduces challenges for general-purpose AV-SSL methods: instruments are frequently visible but silent and multiple similar-looking tools are present in the scene at the same time.

In this work, we present the first investigation of learned audio-visual correspondence for the localisation of surgical activity. Our contributions are threefold:
\begin{enumerate}
  \item We provide the first proof-of-concept for AV-SSL in surgery, demonstrating that audio-visual correspondence can be leveraged to detect and localise active surgical instruments without microphone arrays.
  \item We propose an AV-SSL method built upon SLAVC~\cite{mo2022closer} and explicitly incorporate silence and multi-instrument scenarios through custom loss terms.
  \item We introduce a novel audio-visual dataset of simulated surgery captured in a realistic operating room, providing weak, activity-level labels for image-sound pairs that enable the training of weakly supervised AV-SSL methods.
\end{enumerate}

\section{Related Work}\label{sec:related_work}

In surgery, acoustic signals have mostly been used to infer semantics, with sensors placed close to or in contact with tools and anatomy~\cite{illanes2018novel,seibold2021realtime,seibold2021femoralstem}, or with OR-level microphones for event detection~\cite{fuchtmann2024audio}. Spatial localisation of surgical sounds has only recently been addressed. Beamforming-based sound source localisation relies on an acoustic camera to align acoustic heatmaps with a co-registered optical image~\cite{seibold2024ssl}. This concept was extended to dynamic scenes by fusing beamforming heatmaps with dynamic RGB-D point clouds, yielding a 4D audio-visual representation that spatially and temporally maps surgical actions~\cite{hein2026sound}. To the best of our knowledge, these are the only works on sound source localisation in surgery, and both rely on a phased microphone array and acoustic beamforming. This brings several limitations, such as a bulky and costly setup and lack of real-time capabilities. In addition, beamforming captures only the physical emission point rather than the full instrument, is prone to reflections in reverberant ORs, and treats detection and localisation as separate stages~\cite{hein2026sound}.

In computer vision, AV-SSL replaces physical localisation with learned audio-visual correspondence. 
Models are trained on unlabelled or weakly labelled video to predict which image regions emit the sound~\cite{chen2021localizing}. 
SLAVC~\cite{mo2022closer} showed that earlier methods overfit and fail on negative samples, i.e., silent objects and off-screen sounds, and proposed regularisation strategies and an evaluation protocol that accounts for these cases. 
Methods have been proposed that incorporate silence and noise into training to make detections more robust~\cite{juanola2025learning}.
Furthermore, pretrained vision-language representations~\cite{park2026hearing} or training-free meta-reasoning with multimodal large language models~\cite{Park_2026_CVPR} have been applied to the task of AV-SSL. 
These methods have so far been developed and evaluated exclusively on general domains. 
In the OR, instruments are often visible but inactive, multiple similar tools are present simultaneously, and background noise is substantial. 
It therefore remains unclear whether learned audio-visual correspondence transfers to surgery, which is the question we address in this work.

\section{Methodology}

\begin{figure}[t]
\centering
\makebox[\textwidth][c]{\includegraphics[width=0.9\textwidth]{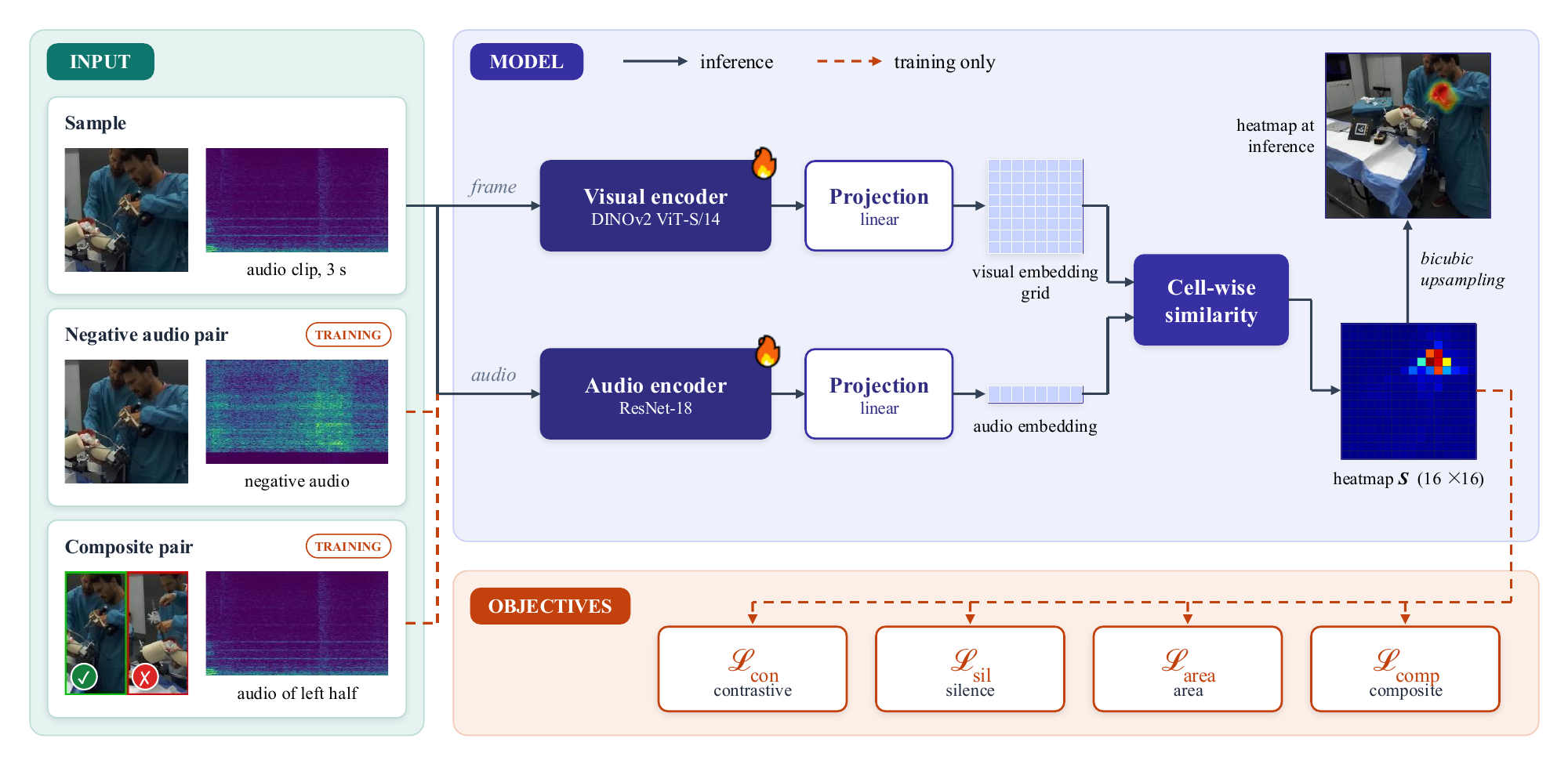}}
\caption{Schematic overview of the method. 
Training adds two types of synthetic samples, each with its own loss term. 
Negative pairs combine a frame with silence or background noise, and composite frames combine the cropped halves of two frames with different instruments. 
The cell-wise cosine similarity between each cell of the visual embedding grid and the audio embedding, averaged over SLAVC's localisation and correspondence spaces, yields the heatmap $S$, which is upsampled at inference time.
}
\label{fig:overview}
\end{figure}

We localise surgical activity in an RGB image from mono audio. 
We fine-tune SLAVC~\cite{mo2022closer}, a weakly supervised sound source localisation model, to predict heatmaps indicating the locations of sound sources. 
We propose to extend the training objective to obtain low-intensity heatmaps for synthetic silence or OR background noise, and enable the model to choose between instruments when more than one is visible. 
\cref{fig:overview} gives an overview of the method and its training objective, and \cref{sec:data} summarises the datasets used for training and evaluation.

\subsection{Model}\label{sec:model}
Our model takes a single RGB frame and a 3\,s mono audio clip and outputs a heatmap over the frame localising the sounding tool or a low-intensity heatmap for silence or background noise. 
It builds on SLAVC~\cite{mo2022closer}, an AV-SSL model with separate encoders for image and audio. 
We replace its visual encoder and extend its contrastive training objective by a silence term, an area term, and a composite term. 
Training requires only one instrument label per recorded sequence and no spatial supervision.

\noindent\textbf{Input.} 
A sample consists of a video frame and the 3\,s of audio centred on it. 
The frame is resized to $448\times448$\,px. 
The audio is resampled to 16\,kHz, scaled to a common RMS level, and converted to a log-power spectrogram. 
At inference, the sample is passed to the model without further processing. 
During training, we additionally augment both modalities. 
Frames are randomly cropped, horizontally flipped, colour-jittered, and with random cutouts applied. 
The audio clips are shifted in time and pitch, scaled in level, and randomly reverberated and low-pass filtered. 
In addition, random frequency bands and time intervals of the spectrogram are masked. 
Unlike these augmentations, which preserve the pairing of a sample, two types of synthetic samples with their own supervision targets are generated during training. 
Negative pairs and composite frames are described with the silence and the composite term below.

\noindent\textbf{Architecture.} 
We build on SLAVC~\cite{mo2022closer}, which embeds audio clips and the grid cells of a frame with two encoders. As heatmap we use SLAVC's inference map $S_{ij}(p) \in [-1,1]$, the cosine 
similarity between audio clip $j$ and grid cell $p$ of frame $i$,
averaged over its localisation and correspondence spaces.
We keep its audio encoder and  projection layers, and change the visual branch, the training objective and use the online encoders at inference. 
The $7\times7$ grid of SLAVC at $224\times224$\,px is too coarse for
instruments that cover only a small part of the frame, so we raise the input resolution to $448\times448$\,px. 
We further replace the ResNet-18 visual encoder by DINOv2 ViT-S/14 with registers~\cite{oquab2024dinov2,darcet2024vision}, which substantially improved the selection of the sounding instrument when multiple instruments are visible (\cref{tab:ablation}).
We average-pool its patch tokens to a $16\times16$ grid, and map them with a linear layer to the input dimension of the SLAVC projection. 
Both encoders start from pretrained weights.

\noindent\textbf{Contrastive term.} 
For training, SLAVC condenses the similarities between frame $i$ and audio clip $j$ into a matching score $M_{ij}$, which is high if one grid cell of the frame matches the clip better than the other cells of the frame and better than the other clips of the batch~\cite{mo2022closer}. 
Only the best cell enters the score, so no spatial annotation is needed. 
All loss terms are defined per training sample and averaged over the batch. 
SLAVC is trained on general-domain video with several hundred sound classes, and treats all other clips of a batch as negatives.
In contrast, surgical data contain only few instrument classes, so many sampled negatives would show the same instrument as the anchor.
We therefore use supervised contrastive learning with the known instrument labels (\cref{sec:data}). 
With $y_i$ the label of sequence $i$ and the positives $\mathcal{P}_i=\{j:y_j=y_i\}$,

\begin{equation}
\mathcal{L}_{\mathrm{con},i}=-\frac{1}{|\mathcal{P}_i|}\sum_{j\in\mathcal{P}_i}
\left[\log\frac{M_{ij}}{\sum_{k}M_{ik}}+\log\frac{M_{ji}}{\sum_{k}M_{ki}}\right].
\end{equation}

\noindent\textbf{Silence and area terms.} 
Our additional terms act directly on the heatmap $S$, so that the map used at inference is itself shaped during training. The contrastive term only ranks pairs within a batch and is invariant to a shift of all similarities, so it cannot express that nothing is sounding. 
Juanola et al.~\cite{juanola2025learning} address this by pairing each image with digital silence and with Gaussian noise and penalising the squared norm of the resulting similarity map. 
We adopt this term and extend the negative audio to background noise recorded in the operating room (\cref{sec:data}). 
In every training step, we draw for each frame $i$ of the batch a clip $n_i$ at random from a bank of Gaussian noise, recorded background noise, and digital silence, compute the heatmap of frame $i$ for this clip, and penalise any non-zero cosine similarity.
\begin{equation}
\mathcal{L}_{\mathrm{sil},i}=\frac{1}{HW}\sum_{p}S_{i n_i}(p)^2,
\qquad
\mathcal{L}_{\mathrm{area},i}=\frac{1}{HW}\sum_{p}\sigma\big(S_{ii}(p)\big).
\label{eq:sil}
\end{equation}
The clips are augmented like the tool audio and attenuated by a random gain of up to $24$\,dB, so that the signal level is no cue. 
In our setting, this term alone did not separate silence from tool activity. 
We therefore pair it with an area term~\eqref{eq:sil}, where $\sigma$ is the logistic function. 
Similar area regularisers have been used to keep the activated region compact~\cite{park2024can}.
Here, however, the term mainly controls the magnitude of the heatmap. 
It pushes the similarity of all cells outside the sounding region towards negative values.
As a result, the heatmap has a large magnitude for tool audio and stays close to zero for silence and noise, while its maximum still marks the source location. 
We exploit this and use the mean squared heatmap value after upsampling,
$a = \frac{1}{|P|}\sum_{x\in P} \hat S_{ii}(x)^2$, where $P$ is the set of pixels, as the score for tool-activity detection.

\noindent\textbf{Composite term.} 
In single-instrument sequences the instrument is the only candidate, so the model can satisfy the contrastive term by highlighting any visible instrument, regardless of the audio. 
To pose the question which of two instruments is sounding, we build composite frames during training.
With probability $0.5$, a training frame in which the instrument lies in the central half of the image is combined with a frame that shows a different instrument. 
The central halves of the two frames are placed side by side or one above the other, in random order.
The audio of the first frame is kept, so that only one of the two visible instruments is sounding.
For each composite sample, the maximum of the heatmap in the half of the sounding instrument, $\Omega_i^{+}$, must exceed the maximum in the other half, $\Omega_i^{-}$, by a margin $m=0.5$,
\begin{equation}
\mathcal{L}_{\mathrm{comp},i}=\max\Big(0,\;m+\max_{p\in\Omega_i^{-}}S_{ii}(p)-\max_{p\in\Omega_i^{+}}S_{ii}(p)\Big).
\end{equation}

\noindent\textbf{Training.} 
The four terms are combined into
\begin{equation}
\mathcal{L}=\mathcal{L}_{\mathrm{con}}+\lambda_{\mathrm{sil}}\mathcal{L}_{\mathrm{sil}}
+\lambda_{\mathrm{area}}\mathcal{L}_{\mathrm{area}}+\lambda_{\mathrm{comp}}\mathcal{L}_{\mathrm{comp}},
\label{eq:loss}
\end{equation}
with $\lambda_{\mathrm{sil}}=0.3$, $\lambda_{\mathrm{area}}=5$, and $\lambda_{\mathrm{comp}}=0.5$, determined empirically. 
We initialise from the published weights of SLAVC for the audio encoder,  DINOv2's weights for the vision encoder and fine-tune the last stage of the audio encoder, the last two blocks of the visual encoder, and the projection layers. All other weights stay frozen. 
The checkpoint with the highest IoU on the validation set is selected. 

\subsection{Data}\label{sec:data}

We train and evaluate the model on two corpora (\cref{tab:data}) recorded in the same operating room (OR-X, Balgrist University Hospital), which contain sequences of simulated surgical drilling, sawing and chiselling on a plastic bone model (SYNBONE AG, Zizers, Switzerland). 
Each sequence shows one active instrument and is labelled with its class, which is the only label used for training.
A new action camera corpus provides the training and validation data, the background noise for the silence term, and test splits for the detection and selection of the sounding instrument.
The public acoustic camera corpus of Hein et al.~\cite{hein2026sound} is used only to test the localisation accuracy.

\noindent\textbf{Action camera corpus.} 
The corpus contains 57 sequences of about one minute each recorded with a GoPro HERO13 Black and a R{\O}DE VideoMic VMMC+, covering three operators, static and hand-held camera viewpoints, and captures in the operating room and an adjacent corridor for variation in the surroundings. 
In the static sequences the camera is centred on the operating site, so that the active instrument lies in the central half of the image, as required by the composite term (\cref{sec:model}). 
Of the 57 sequences, 54 contain a single instrument.
We take $6$ of them for validation and train on the remaining 48.
The 3 test sequences contain a second, idle instrument next to the active one.
Bounding boxes were annotated on the validation and test sequences only and are used for checkpoint selection, threshold calibration and evaluation, never as a training signal. 
Seven additional recordings of the operating room without instrument activity (62\,s, not listed in Table~\ref{tab:data}) provide background noise, five for the silence term and two held out for the detection test.

\noindent\textbf{Acoustic camera corpus.}
The public corpus from \cite{hein2026sound} contains 20 sequences of 20\,s, recorded with a ZED\,2i RGB-D camera and a Ring48 acoustic camera (gfai tech GmbH, Berlin, Germany) while the instruments were tracked with a FusionTrack 500 (Atracsys, Puidoux, Switzerland). 
From these recordings, the model receives only the left RGB image of the stereo camera and the audio of one microphone of the acoustic camera. 
The acoustic heatmaps and the tracking data are not used as model input. 
Ground-truth boxes are only used at test time and obtained by projecting 3D instrument models into the image based on the infrared tracking and drawing a box around the segmentation masks. 

\begin{table}[t]
\caption{Data splits. Bounding box annotations are used for model selection and evaluation only.}\label{tab:data}
\begin{tabular}{@{}llcccl@{}}
\toprule
Split & Corpus & Sequences & Frames & Annotated & Used for \\
\midrule
Training          & action camera   & 48 & 4006 & --        & training \\
Validation       & action camera   & 6  & 371 & $\checkmark$ & validation \\
Two-tool test     & action camera   & 3  & 106 & $\checkmark$       & selection, silence \\
Localisation test & acoustic camera \cite{hein2026sound} & 20 & 853 & $\checkmark$ & localisation \\
\botrule
\end{tabular}
\end{table}


\section{Experiments}
\label{sec:exp}

All metrics are computed on the heatmap of a frame and an audio clip,
upsampled to $224\times224$\,px. We train three models with different seeds
and report mean and standard deviation.

\noindent\textbf{Localisation.} 
We evaluate the model's instrument localisation accuracy on the acoustic camera corpus \cite{hein2026sound} and report the results in \cref{tab:loc}.
For this purpose, we convert each heatmap prediction into a bounding box.
The heatmap is normalised to $[0,1]$ by its own minimum and maximum, and pixels above a threshold form the predicted region.
The bounding box around its largest connected component is compared to the ground truth by intersection over union (IoU). 
The threshold is determined empirically on the validation set for each model separately, including the baselines and the ablations. 
In addition, we report the median distance in pixels between the maximum of the heatmap and the centre of the ground-truth box, at a resolution of $224\times224$\,px. 
We compare against SLAVC~\cite{mo2022closer} and ACL-SSL~\cite{park2024can} with their published weights, and against SLAVC fine-tuned on our data at 448 px with the authors' objective and regularisation, using the instrument class as positive and the maximum over cells as in our model, with the same data, schedule and checkpoint selection. Its heatmaps are read from the online encoders.

\begin{table}[h]
\setlength{\tabcolsep}{4pt}
\caption{Localisation on the acoustic camera corpus.
IoU and median distance between the heatmap maximum and the centre of the ground-truth box in px at $224\times224$, per instrument and over all frames. 
SLAVC (fine-tuned) is fine-tuned on our data with the authors'
regularisation and class positives. 
Fine-tuned models: mean $\pm$ standard deviation over three seeds.}
\label{tab:loc}
\begin{tabular}{@{}lcccccccc@{}}
\toprule
& \multicolumn{4}{c}{IoU} & \multicolumn{4}{c}{Distance [px]} \\
\cmidrule(lr){2-5}\cmidrule(lr){6-9}
Model & Saw & Chisel & Drill & All & Saw & Chisel & Drill & All \\
\midrule
ACL-SSL~\cite{park2024can} & 0.04 & 0.02 & 0.02 & 0.03 & 47 & 28 & 77 & 37 \\
SLAVC~\cite{mo2022closer}  & 0.05 & 0.01 & 0.06 & 0.04 & 62 & 74 & 70 & 67 \\
\midrule
SLAVC (fine-tuned) & $0.28\pm0.03$ & $0.12\pm0.05$ & $0.30\pm0.06$ & $0.23\pm0.04$ & $10\pm2$ & $17\pm5$ & $12\pm1$ & $14\pm4$ \\
Ours & $0.46\pm0.04$ & $0.14\pm0.02$ & $0.50\pm0.04$ & $0.36\pm0.03$ & $9\pm1$ & $23\pm8$ & $12\pm1$ & $12\pm2$ \\
\botrule
\end{tabular}
\end{table}

\noindent\textbf{Detection.} 
We test the activity score, the mean squared value of the heatmap, on the two-tool test split. 
Each frame is paired with its own audio as positive and with each of $17$ overlapping windows of $3\,s$ ($1\,s$ hop) of operating room background noise as negatives.
We report the area under the ROC curve (AUC), which rates the ranking of the scores and does not depend on a threshold. 
The score separates tool activity from background noise completely, with an AUC of $1.00$ for all three seeds and all three instruments. 
The lowest score of a tool frame lies between $0.60$ and $0.73$ and the highest score of a background clip between $0.41$ and $0.51$, depending on the seed. SLAVC fine-tuned achieves an AUC of $0.48$ on our activity score.

\begin{figure}[t]
\centering
\includegraphics[width=0.32\textwidth, keepaspectratio]{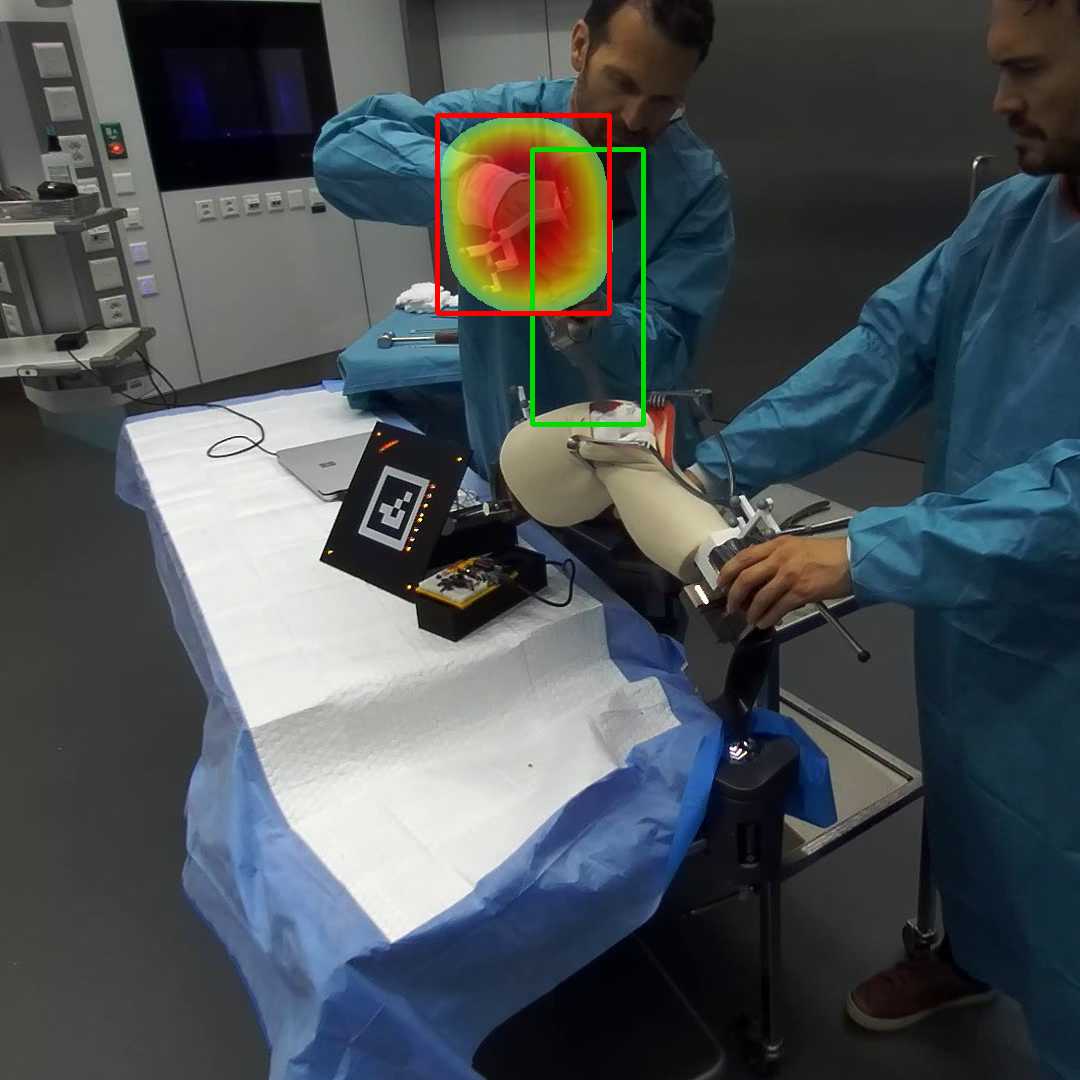}\hfill
\includegraphics[width=0.32\textwidth, keepaspectratio]{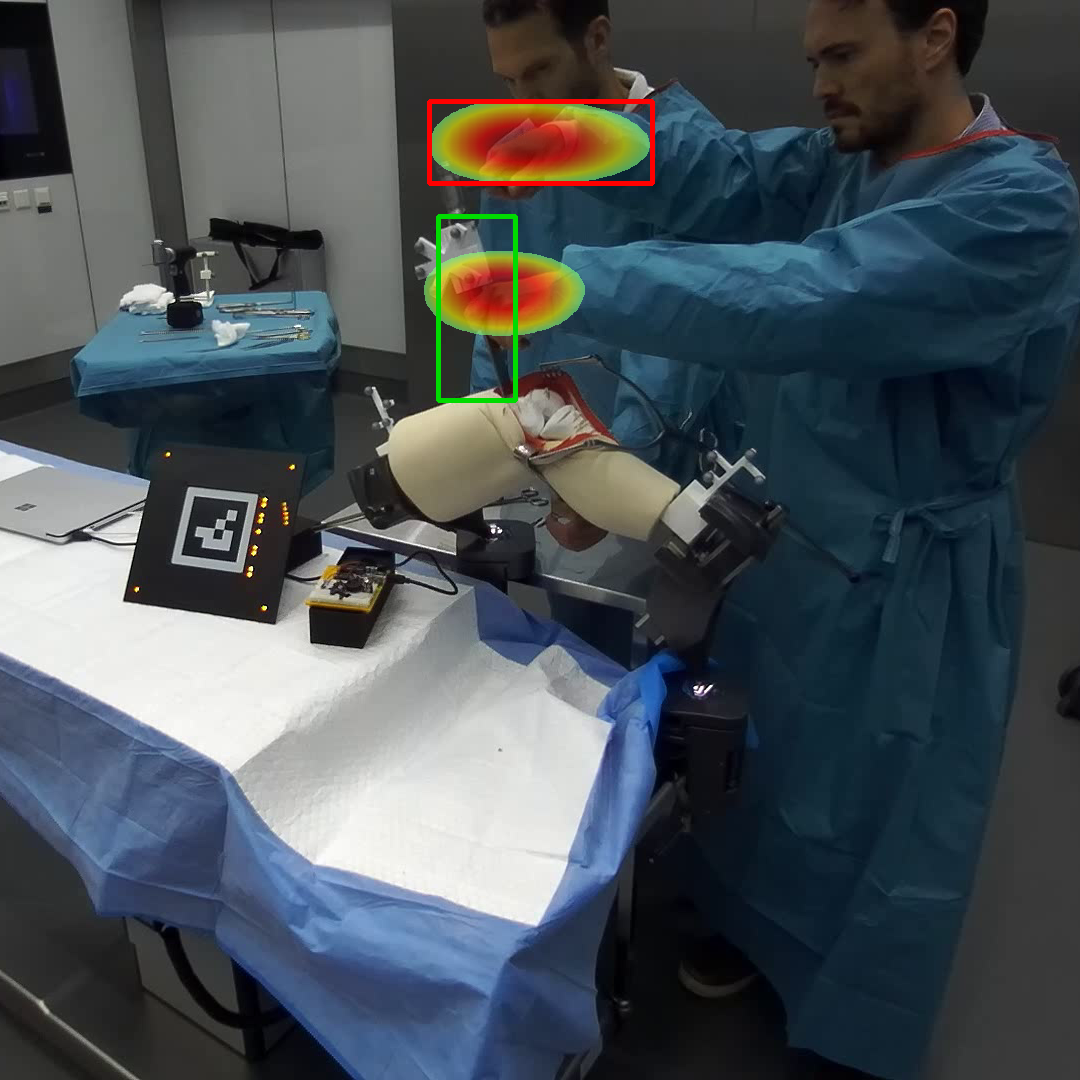}\hfill
\includegraphics[width=0.32\textwidth, keepaspectratio]{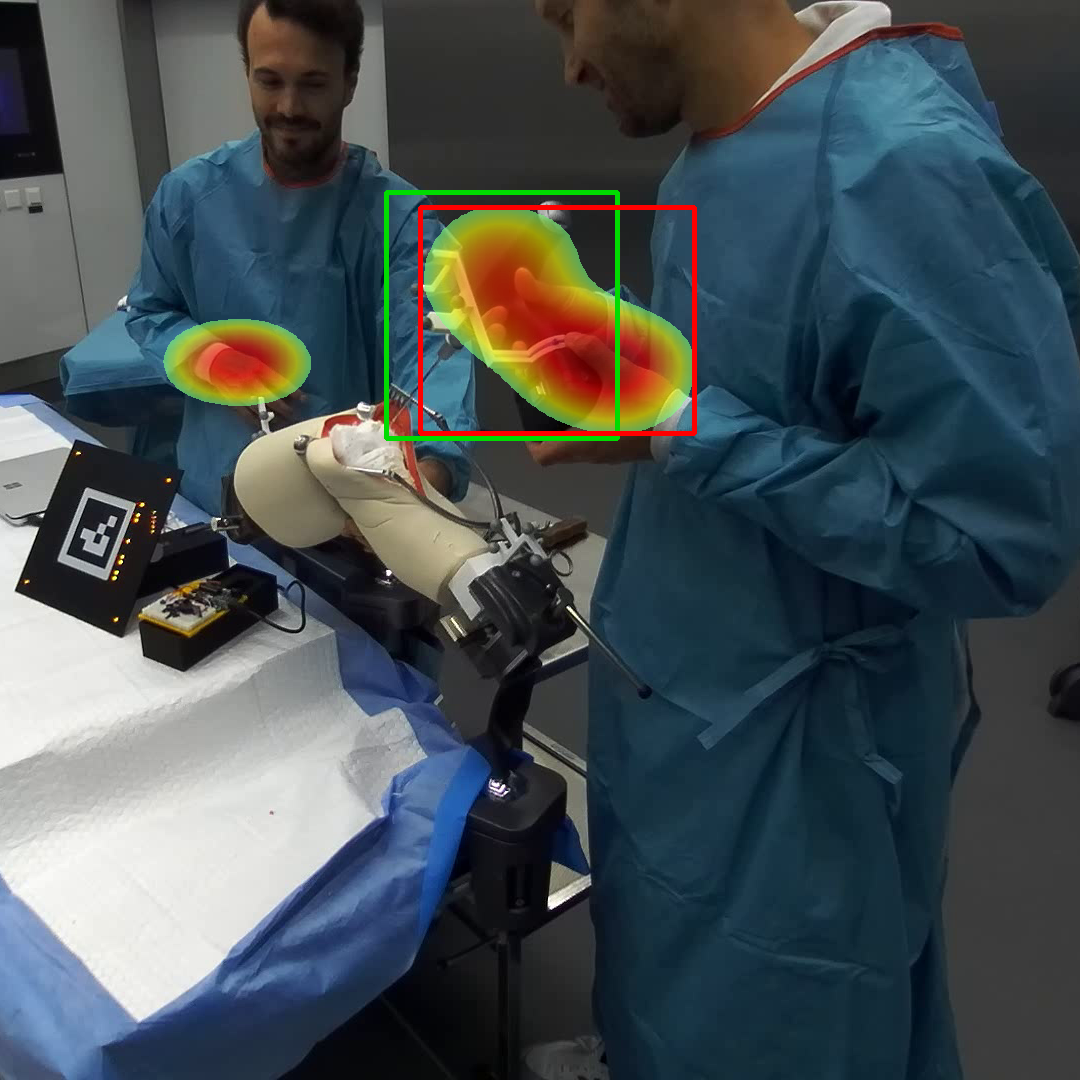}\\[2pt]

\caption{Heatmaps of the proposed model on the acoustic camera corpus.
Heatmaps are shown above half of their range. Green: ground-truth box. Red: prediction bounding box. Left: saw. Middle: chisel, where the model
responds to the hand swinging the hammer as well as to the chisel
(\cref{sec:dis}). Right: drill, where the empty hand of the assistant
also receives part of the response.}
\label{fig:qual}
\end{figure}

\begin{figure}[t]
\centering
\includegraphics[width=0.32\textwidth]{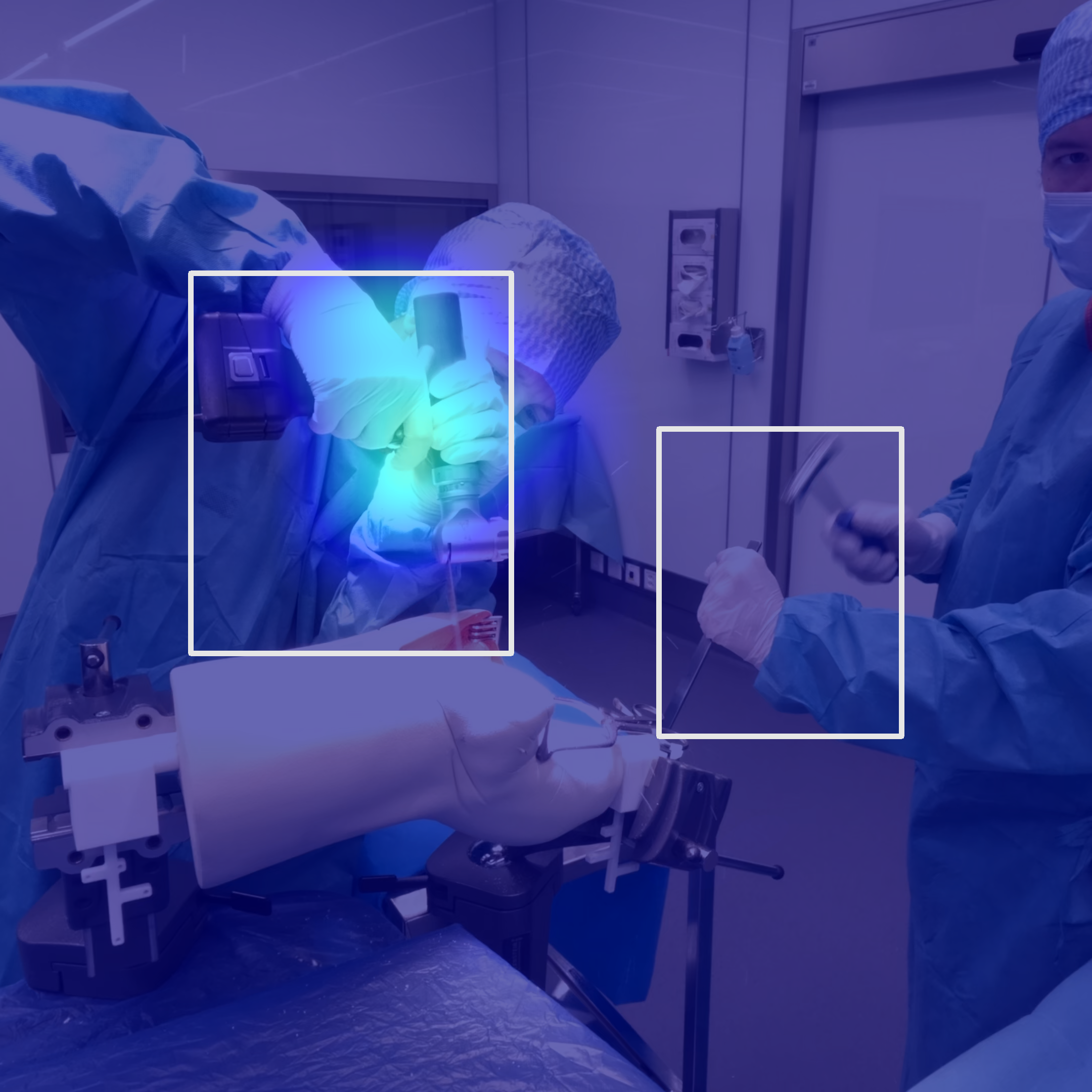}\hfill
\includegraphics[width=0.32\textwidth]{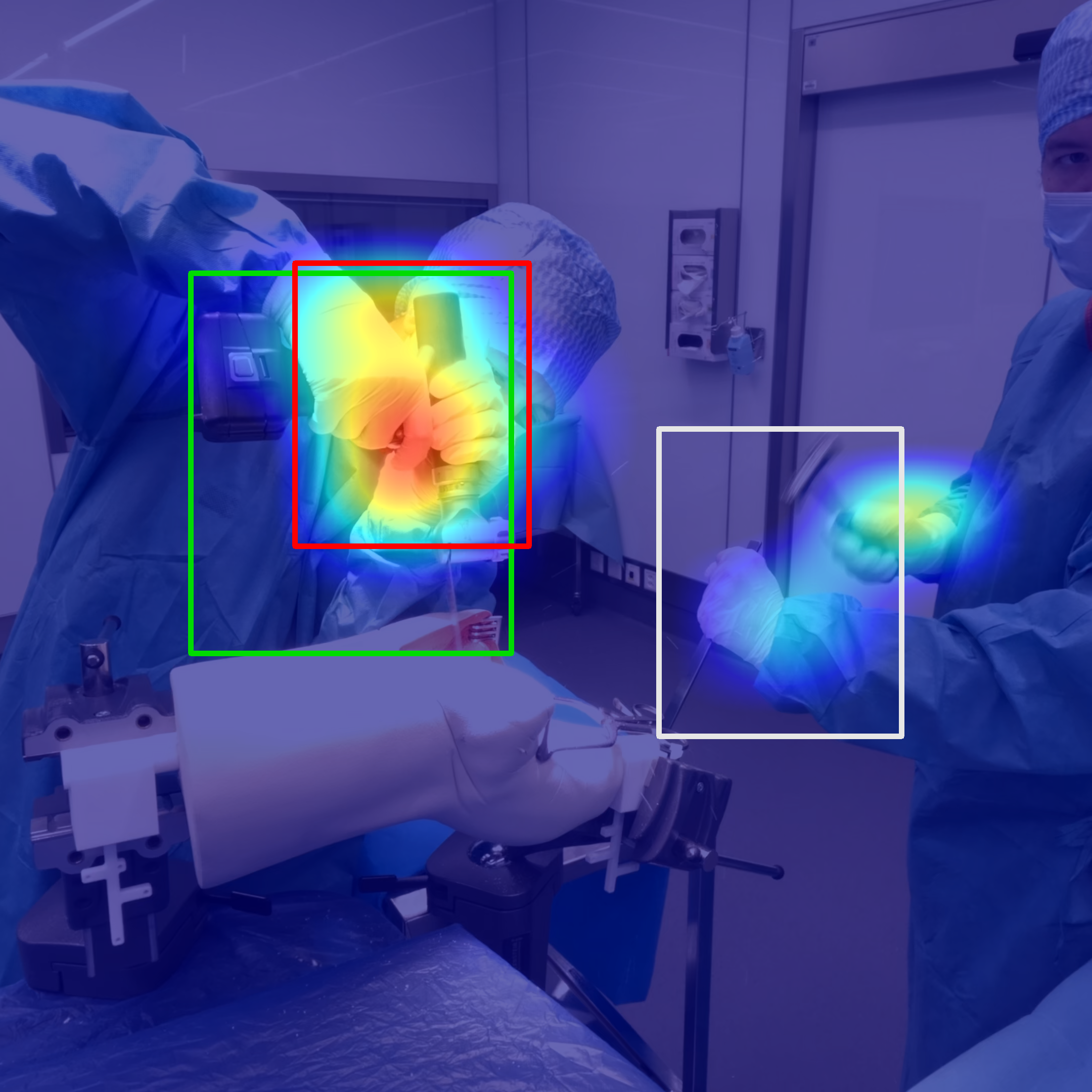}\hfill
\includegraphics[width=0.32\textwidth]{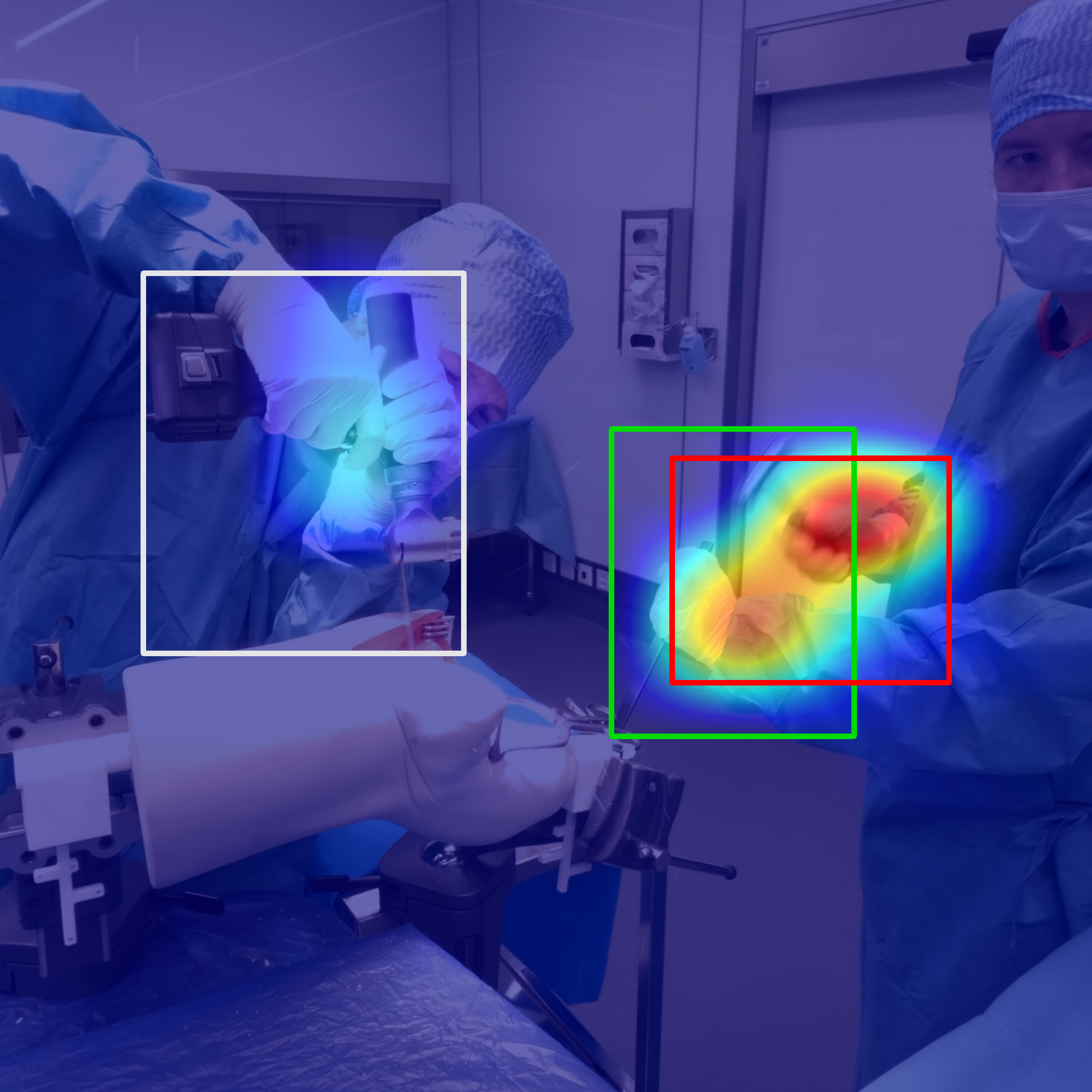}
\caption{One frame of the two-tool test split under different audio. Green: GT of correct instrument. Grey: GT of wrong instrument. Red: predicted bounding box. Background noise classified as inactive
(left), matched saw audio (middle), and swapped to chisel audio (right). For visualisation purposes, each heatmap is shown as the deviation from its mean, on one common colour scale.}
\label{fig:twotool}
\end{figure}

\noindent\textbf{Tool selection.}
Each frame of the two-tool test split (\cref{fig:twotool}) is evaluated with its own audio (matched) and with audio of the idle instrument's class (swapped). 
A prediction is correct if the maximum of the heatmap inside the box of the instrument that matches the audio exceeds that inside the other box. 
We report the accuracy for both conditions. 
With the frame's own audio, the active instrument is correctly recovered with an accuracy of $0.93\pm0.05$. 
With the audio of the idle instrument's class, the response moves to the idle instrument with an accuracy of $0.87 \pm 0.10$. SLAVC fine-tuned achieves $0.81\pm0.14$ matched and $0.80\pm0.05$ swapped (mean and standard deviation over three seeds). 

\noindent\textbf{Ablations.} 
\cref{tab:ablation} analyses the contributions of each training loss and the choice of the visual encoder. 
The silence term and the area term are required for detection (AUC) and the composite term for instrument selection (matched and swapped). The ablation row without area term has the best selection accuracy, which shows that the area term trades part of the selection margin for reliable detection.
The ResNet-18 vision encoder shows similar IoU, but is weaker on selecting between two instruments. 

\begin{table}[t]
\setlength{\tabcolsep}{4pt}
\caption{Ablation of the loss terms and the visual encoder, evaluated on the two-tool test split. The contrastive term is always used. First row: the proposed model as in \cref{tab:loc}.
All rows are means over three seeds.}
\label{tab:ablation}
\begin{tabular}{@{}cccl ccccc@{}}
\toprule
\multicolumn{3}{c}{} & & \multicolumn{2}{c}{Localisation} & Detection & \multicolumn{2}{c}{Selection} \\
 \cmidrule(lr){5-6} \cmidrule(lr){7-7} \cmidrule(l){8-9}
$\mathcal{L}_{\mathrm{sil}}$ & $\mathcal{L}_{\mathrm{area}}$ & $\mathcal{L}_{\mathrm{comp}}$ & Visual encoder
& IoU & Median Dist. px & AUC & Matched & Swapped \\
\midrule
\checkmark & \checkmark & \checkmark & DINOv2    & \textbf{0.36} & 11.7 & \textbf{1.00} & 0.93 & 0.87 \\
\midrule
\checkmark & \checkmark & \checkmark & ResNet-18 & 0.32 & \textbf{9.3} & \textbf{1.00} & 0.78 & 0.73 \\
\checkmark & \checkmark & $\times$   & DINOv2    & 0.25 & 23.5 & \textbf{1.00} & 0.48 & 0.62 \\
\checkmark & $\times$   & \checkmark & DINOv2    & 0.32 & 10.9 & 0.61 & \textbf{0.97} & \textbf{0.98} \\
$\times$   & \checkmark & \checkmark & DINOv2    & 0.34 & 11.0 & 0.06 & 0.86 & 0.92 \\
$\times$   & $\times$   & $\times$   & DINOv2    & 0.16 & 20.6 & 0.39 & 0.58 & 0.63 \\

\botrule
\end{tabular}
\end{table}

\noindent\textbf{Runtime.} 
Measured on a Quadro RTX 6000 with a batch size of $1$, a forward pass of our method takes about $58\,ms$ ($\approx$17\,fps). 
Because each prediction uses a $3\,\mathrm{s}$ audio window centred on the frame, the model needs $1.5\,\mathrm{s}$ of future audio, which brings the end-to-end latency to about $1.56\,\mathrm{s}$.

\section{Discussion}
\label{sec:dis}

This paper presents a novel approach that enables the detection and localisation of surgical tool activity in one model at a low cost, leveraging the natural co-occurrence of video and audio as a training signal. 
Trained on about an hour of video from an action camera with a single microphone, labelled only with the instrument per sequence, the model localises the sounding instrument on an unseen corpus recorded with different cameras and microphones. 
The spatial information, the expensive part of manual annotation, is learned from the correspondence. The instrument labels are cheap to annotate and could be replaced by an audio classifier in future
work~\cite{hein2026sound}. 
We show that the tested zero-shot baselines, pretrained on general-domain video recordings, do not transfer to the operating room setting, requiring in-domain learning. SLAVC fine-tuned, shows promising localisation and selection but struggles with the detection of OR background noise. The proposed adapted architecture outperforms the reference model in detection, localisation, and selection.

Our evaluation shows that the saw and drill can be localised with a median distance of about $10$\,px to the centre of the ground-truth box, about $5\,\%$ of the frame width. 
The less accurate results for surgical chiselling are partly a matter of ground truth, which does not include annotations of the hammer, so responses on the hammer reduce localisation scores.
Our model associates chiselling sounds with the chisel and hammer (\cref{fig:qual}), arguably correctly, as their collision produces the sound.

For our approach, the composite term supports the conditioning of the model on audio. The selection test demonstrates that the model then makes this decision based on the audio. Without it, selection accuracy drops to chance in the matched condition, as illustrated in our ablation study.  


\noindent\textbf{Limitations.}
The diversity of our training and evaluation data is limited, covering three orthopaedic instruments, simulated actions on bone models, and a single operating room.
Localisation was evaluated on a public corpus recorded from a single viewpoint, although the two-tool test split, recorded from other viewpoints, indicates that selection does not depend on a specific view.
The recordings contain no other OR sounds, such as speech or monitor alarms, and detection was tested only against background noise.
Recordings of real procedures in different ORs are needed to test robustness to such sounds and to quieter actions such as coagulation or suction in visceral surgery.
However, our initial results motivate further investigation of audio-visual correspondence learning for general surgical scene understanding. 

Training requires sequence-level instrument labels, and the composite term relies on frames with the active instrument in the central half of the image.
The output is still coarse in space and time, as a $16\times16$ heatmap only indicates the region of the sounding instrument, and each prediction uses 3\,s of audio centred on the frame.
Finer feature grids and shorter audio windows could extend this towards a spatio-temporal segmentation of the sounding instrument.
Since hand-held instruments always appear with hands, hands act as a confounder, and empty hands can receive part of the response (\cref{fig:qual}, right).
Training frames with empty hands as hard negatives may reduce this.
Finally, the model assumes that the sounding instrument is visible.

\section{Conclusion}
We present the first proof-of-concept for sounding object detection based on audio-visual correspondence in the surgical domain. 
Our method localises and disambiguates tools reliably, is trained on data recorded with commodity hardware without localisation supervision, and enables the detection and localisation of surgical activity for advanced multimodal surgical scene understanding. 


\noindent\textbf{Data, code and/or material availability}
The code and dataset will be released upon acceptance.



\bibliography{sn-bibliography}

@article{Mezger2013,
  author  = {Mezger, Uli and Jendrewski, Claudia and Bartels, Michael},
  title   = {Navigation in surgery},
  journal = {Langenbeck's Archives of Surgery},
  year    = {2013},
  volume  = {398},
  number  = {4},
  pages   = {501--514}
}

@article{hein2017sds,
author={Maier-Hein, Lena      and
Vedula, Swaroop S.    and
Speidel, Stefanie     and
Navab, Nassir         and
Kikinis, Ron          and
Park, Adrian          and
Eisenmann, Matthias   and
Feussner, Hubertus    and
Forestier, Germain    and
Giannarou, Stamatia   and
Hashizume, Makoto     and
Katic, Darko          and
Kenngott, Hannes      and
Kranzfelder, Michael  and
Malpani, Anand        and
März, Keno            and
Neumuth, Thomas       and
Padoy, Nicolas        and
Pugh, Carla           and
Schoch, Nicolai       and
Stoyanov, Danail      and
Taylor, Russell       and
Wagner, Martin        and
Hager, Gregory D.     and
Jannin, Pierre},
title={Surgical data science for next-generation interventions},
journal={Nature Biomedical Engineering},
volume=1,
pages={691--696},
year=2017
}

@article{oquab2024dinov2,
  title   = {{DINOv2}: Learning Robust Visual Features without Supervision},
  author  = {Oquab, Maxime and Darcet, Timoth{\'e}e and Moutakanni, Th{\'e}o and
             Vo, Huy and Szafraniec, Marc and Khalidov, Vasil and
             Fernandez, Pierre and Haziza, Daniel and Massa, Francisco and
             El-Nouby, Alaaeldin and Assran, Mahmoud and Ballas, Nicolas and
             Galuba, Wojciech and Howes, Russell and Huang, Po-Yao and
             Li, Shang-Wen and Misra, Ishan and Rabbat, Michael and
             Sharma, Vasu and Synnaeve, Gabriel and Xu, Hu and
             J{\'e}gou, Herv{\'e} and Mairal, Julien and Labatut, Patrick and
             Joulin, Armand and Bojanowski, Piotr},
  journal = {Transactions on Machine Learning Research},
  year    = {2024}
}

@inproceedings{juanola2025learning,
author    = {Xavier Juanola and Giovana Morais and Magdalena Fuentes and Gloria Haro},
title     = {Learning from Silence and Noise for Visual Sound Source Localization},
booktitle = {36th British Machine Vision Conference 2025},
organization = {BMVA},
year      = {2025}
}

@inproceedings{darcet2024vision,
  title={Vision transformers need registers},
  author={Darcet, Timoth{\'e}e and Oquab, Maxime and Mairal, Julien and Bojanowski, Piotr},
  booktitle={International conference on learning representations},
  volume={2024},
  pages={2632--2652},
  year={2024}
}

@inproceedings{park2024can,
  title={Can clip help sound source localization?},
  author={Park, Sooyoung and Senocak, Arda and Chung, Joon Son},
  booktitle={2024 IEEE/CVF Winter Conference on Applications of Computer Vision (WACV)},
  pages={5699--5708},
  year={2024},
  organization={IEEE}
}

@ARTICLE{twinanda2017endonet,
  author={Twinanda, Andru P. and Shehata, Sherif and Mutter, Didier and Marescaux, Jacques and de Mathelin, Michel and Padoy, Nicolas},
  journal={IEEE Transactions on Medical Imaging}, 
  title={EndoNet: A Deep Architecture for Recognition Tasks on Laparoscopic Videos}, 
  year={2017},
  volume={36},
  number={1},
  pages={86--97}}

@InProceedings{oeszoy2022ssg,
author="{\"O}zsoy, Ege
and {\"O}rnek, Evin P{\i}nar
and Eck, Ulrich
and Czempiel, Tobias
and Tombari, Federico
and Navab, Nassir",
editor="Wang, Linwei
and Dou, Qi
and Fletcher, P. Thomas
and Speidel, Stefanie
and Li, Shuo",
title="4D-OR: Semantic Scene Graphs for OR Domain Modeling",
booktitle="Medical Image Computing and Computer Assisted Intervention -- MICCAI 2022",
year="2022",
pages="475--485"
}

@Article{ding2024dt,
AUTHOR = {Hao Ding and Lalithkumar Seenivasan and Benjamin D. Killeen and Sue Min Cho and Mathias Unberath},
TITLE = {Digital twins as a unifying framework for surgical data science: the enabling role of geometric scene understanding},
JOURNAL = {Artificial Intelligence Surgery},
VOLUME = {4},
YEAR = {2024},
NUMBER = {3}
}

@article{illanes2018novel,
author = {Illanes, Alfredo  and Boese, Axel and Maldonado, Ivan and Pashazadeh, Ali and Schaufler, Anna and Navab, Nassir and Friebe, Michael},
title = {Novel clinical device tracking and tissue event characterization using proximally placed audio signal acquisition and processing},
journal = {Scientific Reports},
year = {2018},
volume = {8}
}

@article{seibold2021realtime,
title = {Real-time acoustic sensing and artificial intelligence for error prevention in orthopedic surgery},
author = {Seibold, Matthias and Maurer, Steven and Hoch, Armando and Zingg, Patrick and Farshad, Mazda and Navab, Nassir and Fürnstahl, Philipp},
journal = {Scientific Reports},
year = {2021},
volume = {11},
issue = {3993}
}

@inproceedings{seibold2021femoralstem,
title = {Acoustic-Based Spatio-Temporal Learning for Press-Fit Evaluation of Femoral Stem Implants},
author = {Seibold, Matthias and Hoch, Armando and Suter, Daniel and Farshad, Mazda and Zingg, Patrick O. and Navab, Nassir and Fürnstahl, Philipp},
booktitle = {International Conference on Medical Image Computing and Computer-Assisted Intervention},
year = {2021},
pages = {447--456}
}

@article{fuchtmann2024audio,
  title={Audio-based event detection in the operating room},
  author={Fuchtmann, Jonas and Riedel, Thomas and Berlet, Maximilian and Jell, Alissa and Wegener, Luca and Wagner, Lars and Graf, Simone and Wilhelm, Dirk and Ostler-Mildner, Daniel},
  journal={International Journal of Computer Assisted Radiology and Surgery},
  volume={19},
  number={12},
  pages={2381--2387},
  year={2024},
  publisher={Springer}
}

@InProceedings{seibold2024ssl,
    author    = {Seibold, Matthias and Bahari Malayeri, Ali and Fürnstahl, Philipp},
    title     = {Spatial Context Awareness in Surgery Through Sound Source Localization},
    booktitle = {International Conference on Medical Image Computing and Computer-Assisted Intervention (MICCAI)},
    year      = {2024}
}

@article{hein2026sound,
  title={Sound source localization for spatial mapping of surgical actions in dynamic scenes},
  author={Hein, Jonas and Vlachopoulos, Lazaros and Olthof, Maurits Geert Laurent and Sigrist, Bastian and Fürnstahl, Philipp and Seibold, Matthias},
  journal={International Journal of Computer Assisted Radiology and Surgery},
  pages={1--9},
  year={2026},
  publisher={Springer}
}

@inproceedings{chen2021localizing,
  title={Localizing visual sounds the hard way},
  author={Chen, Honglie and Xie, Weidi and Afouras, Triantafyllos and Nagrani, Arsha and Vedaldi, Andrea and Zisserman, Andrew},
  booktitle={2021 IEEE/CVF Conference on Computer Vision and Pattern Recognition (CVPR)},
  pages={16862--16871},
  year={2021},
  organization={IEEE}
}

@article{mo2022closer,
  title={A closer look at weakly-supervised audio-visual source localization},
  author={Mo, Shentong and Morgado, Pedro},
  journal={Advances in Neural Information Processing Systems},
  volume={35},
  pages={37524--37536},
  year={2022}
}

@article{park2026hearing,
  title={Hearing and seeing through clip: A framework for self-supervised sound source localization},
  author={Park, Sooyoung and Senocak, Arda and Chung, Joon Son},
  journal={International Journal of Computer Vision},
  volume={134},
  number={4},
  pages={179},
  year={2026},
  publisher={Springer}
}

@InProceedings{Park_2026_CVPR,
    author    = {Park, Subin and Kim, Jung Uk},
    title     = {Generate, Analyze, and Refine: Training-Free Sound Source Localization via MLLM Meta-Reasoning},
    booktitle = {Proceedings of the IEEE/CVF Conference on Computer Vision and Pattern Recognition (CVPR)},
    month     = {June},
    year      = {2026},
    pages     = {15752--15761}
}

\end{document}